\documentclass[11pt]{article}

\usepackage[cp1251]{inputenc}
\usepackage[english]{babel}
\usepackage{abstract}

\usepackage{amsmath}
\usepackage{amssymb}
\usepackage{euscript}
\usepackage{hyperref}

\usepackage{epsfig, graphicx}
\usepackage{caption}

\makeatletter
\renewcommand{\@oddhead}{\small Published in Petrology, 2016, Vol. 24, No. 1, pp. 21-34. \hfill}
\renewcommand{\@oddfoot}{\hfill ---~\thepage~---\hfill}
\makeatother

\begin{document}

\begin{center}
\Large Production of $^3$He in Rocks by Reactions Induced by Particles of the Nuclear-Active and Muon Components of Cosmic Rays: Geological and Petrological Implications

\vspace{0.5cm}
\large A.V. Nesterenok$^1$ and O.V. Yakubovich$^{2, 3}$
\vspace{0.5cm}
\end{center}

\noindent
$^1$ \normalsize Ioffe Physical-Technical Institute, Russian Academy of Sciences, Polytechnicheskaya St. 26, St.~Petersburg, 194021 Russia

\noindent
e-mail: alex-n10@yandex.ru

\noindent
$^2$ Institute of Precambrian Geology and Geochronology, Russian Academy of Sciences, nab. Makarova 2, St. Petersburg, 199034 Russia

\noindent
e-mail: olya.v.yakubovich@gmail.com

\noindent
$^3$ St. Petersburg State University, Universitetskaya nab. 7/9, St. Petersburg, 199034 Russia

\begin{abstract} 
\noindent
The paper presents data on the production of the $^3$He nuclide in rocks under the effect of cosmic-ray particles. The origin of the nuclide in the ground in neutron- and proton-induced spallation reactions, reactions induced by high-energy muons, and negative muon capture reactions is analyzed. The cross sections of reactions producing $^3$He and $^3$H are calculated by means of numerical simulations with the GEANT4 simulation toolkit. The production rate of the $^3$He nuclide in the ground is evaluated for the average level of solar activity at high geomagnetic latitudes and at sea level. It is proved that the production of $^3$He in near-surface ground layers by spallation reactions induced by cosmic-ray protons may be approximately 10\% of the total production rate of cosmogenic $^3$He. At depths of 10-50 m.w.e., the accumulation of $^3$He is significantly contributed by reactions induced by cosmic-ray muons. Data presented in the paper make it possible to calculate the accumulation rate of $^3$He in a rock depending on depth that is necessary for the evaluation of the exposure time of the magmatic or metamorphic complex on the Earth's surface ($^3$He dating).
\end{abstract}

\section{Introduction}
\noindent
Cosmogenic nuclides provide a principally important tool for studying processes on the Earth's surface. These nuclides are produced in situ in rocks composing the Earth's surface by reactions induced by cosmic radiation. Currently cosmogenic nuclides ($^{10}$Be, $^{14}$C, $^{26}$Al, $^{36}$Cl, and others) are utilized to solve certain problems of geomorphology, glaciology, soil science, palaeoclimatology, and volcanology (Lal, 1991; Bierman, 1994; Gosse and Phillips, 2001; Dunai, 2010). Knowing the content of a nuclide in a mineral and the production rate of this nuclide, one can calculate the surface exposure time of the rock and determine when the magmatic or metamorphic complex was brought to the modern erosion surface.

A cosmogenic nuclide can be utilized as a geochronometer only if the concentration of this nuclide in the rock is very low, and hence, the cosmogenic component can thus be discernible compared to the component captured by the rock during its crystallization. For most rock-forming minerals, these nuclides are stable isotopes of noble gases whose concentrations are very low, such as $^{21}$Ne, $^{22}$Ne, $^{36}$Ar, $^{38}$Ar, etc., including the He isotope $^3$He (Porcelli et al., 2002; Dunai, 2010). Although the content of $^3$He on the Earth is very low, this isotope has long not been used for the purposes of cosmogenic geochronology (Farley et al., 2006) because He can easily migrate from the crystal lattices of rock-forming minerals (such as quartz and feldspars) even at room temperature (Shuster and Farley, 2005). The only exception is large 'perfect' quartz crystals of hydrothermal and/or metasomatic genesis, whose rates of He loss are relatively low because of the large size of the diffusion cells (Tolstikhin et al., 1974). Because of this, systems based on the cosmogenic $^3$He isotope have been of little use until recently in evaluating the exposure tiles of rocks, except only for olivine and pyroxene xenocrysts in young lava sheets (Niedermann, 2002). The involvement of such mineral phases as garnet, magnetite, zircon, apatite, and titanite, which preserve much He, in $^3$He dating led several researchers to revise the outlooks of this technique (Gayer et al., 2004; Kober et al., 2005; Farley et al., 2006). This made it principally possible to evaluate the exposure time of magmatic and metamorphic complexes that were brought to the surface relatively long ago and can not hence be studied using radionuclides (Margerison et al., 2005; Evenstar et al., 2009).

To utilize a nuclide in solving problems of geological interest, it is necessary to understand in detail the mechanisms producing the nuclide and to estimate the accumulation rate of this nuclide in the rock. The principal approaches to solving this problem are as follows: numerical simulations underlain by physical principles, irradiation experiments, and obtaining empirical information on concentrations of the nuclide in samples of naturally occurring rocks (Dunai, 2010). Numerical calculations of the accumulation rate of cosmogenic He in rocks were reported in (Lal, 1987; Masarik and Reedy, 1995). Since that time physical models have been remarkably revised and modified, and extensive experimental data were obtained on the fluxes of cosmic-ray particles and the cross sections of nuclear reactions.

The primary goal of our research was to analyse in detail the mechanism producing cosmogenic $^3$He (including its precursor $^3$H) in rocks. We have calculated the accumulation rate of $^3$He in grounds due to reactions triggered by the nuclear-active and muon components of cosmic radiation. We were the first to quantify the production rate of $^3$He in spallation reactions induced by cosmic-ray protons. Also, we have analysed $^3$H production in the thermal neutron capture reaction by Li. We have also evaluated the contributions of various reactions to the integral $^3$He accumulation rate depending on depth in the ground (metamorphic or magmatic complex).

\section{Sources of Cosmogenic Helium}
\noindent
The major source of cosmogenic He on the Earth's surface is the spallation reaction induced by high-energy neutrons of cosmic rays (Dunai, 2010). Cosmic-ray high-energy neutrons cannot enter the ground for any significant depth, and hence, cosmogenic He is produced mostly within a near-surface layer no thicker than 2 m. In studying $^3$He production in rocks, one should also take into account the origin of the $^3$H nuclide: the $\beta$-decay time of $^3$H into $^3$He is approximately 12.3 years, which is a negligibly short time span compared to the surface exposure times of rocks of interest. Another possible source of $^3$He in minerals is reactions induced by high-energy muons of cosmic rays and $\mu^{-}$ capture reactions by atoms (Lal, 1987). Having a smaller interaction cross section than that of particles of the nuclear-active component of cosmic rays (neutrons and protons), muons are able to reach depths as great as dozens of meters. The contribution of the muogenic component to the total $^3$He concentration was usually neglected (Kurz, 1986; Farley et al., 2006).

In rocks rich in Li, much $^3$H (and hence, also $^3$He) is produced by the thermal neutron capture reaction by the nuclei of Li atoms (Mamyrin and Tolstikhin, 1981). A source of thermal neutrons in grounds are the nuclear-active and muon components of cosmic rays (for the ground near the surface) and ($\alpha$, n) reactions, and reactions of U spontaneous fission. Also, rocks can contain certain amounts of primary $^3$He that was captured when the rocks were formed and is commonly contained in gas and gas-liquid inclusions in minerals (Mamyrin and Tolstikhin, 1981; Porcelli et al., 2002). Accumulation of $^3$He in a rock is schematically represented in Figure \ref{fig1a}.

Nowadays extensive experimental information is available on the concentration of cosmogenic $^3$He in rocks from various localities worldwide, and the empirically determined production rate of cosmogenic $^3$He in pyroxene and olivine crystals in near-surface ground layers lies within the range of 110 to 150 atoms per year per gram of material at sea level and at high geomagnetic latitudes (Amidon et al., 2009; Goehring et al., 2010; Licciardi et al., 1999, 2006). The possible reason for this spread is the inaccuracy of extrapolation models currently adopted for $^3$He production at various altitudes and latitudes (Amidon et al., 2008; Gayer et al., 2004; Goehring et al., 2010) and also underestimation of the contribution of $^3$He produced by thermal neutron capture by Li (Dunai et al., 2007). Other possible reasons for this spread of the experimental values of the $^3$He production rate are discussed in (Farley et al., 2006; Blard et al., 2006; Blard and Farley, 2008).

\begin{figure}[t]
	\centering
	\includegraphics[width = 0.7\textwidth]{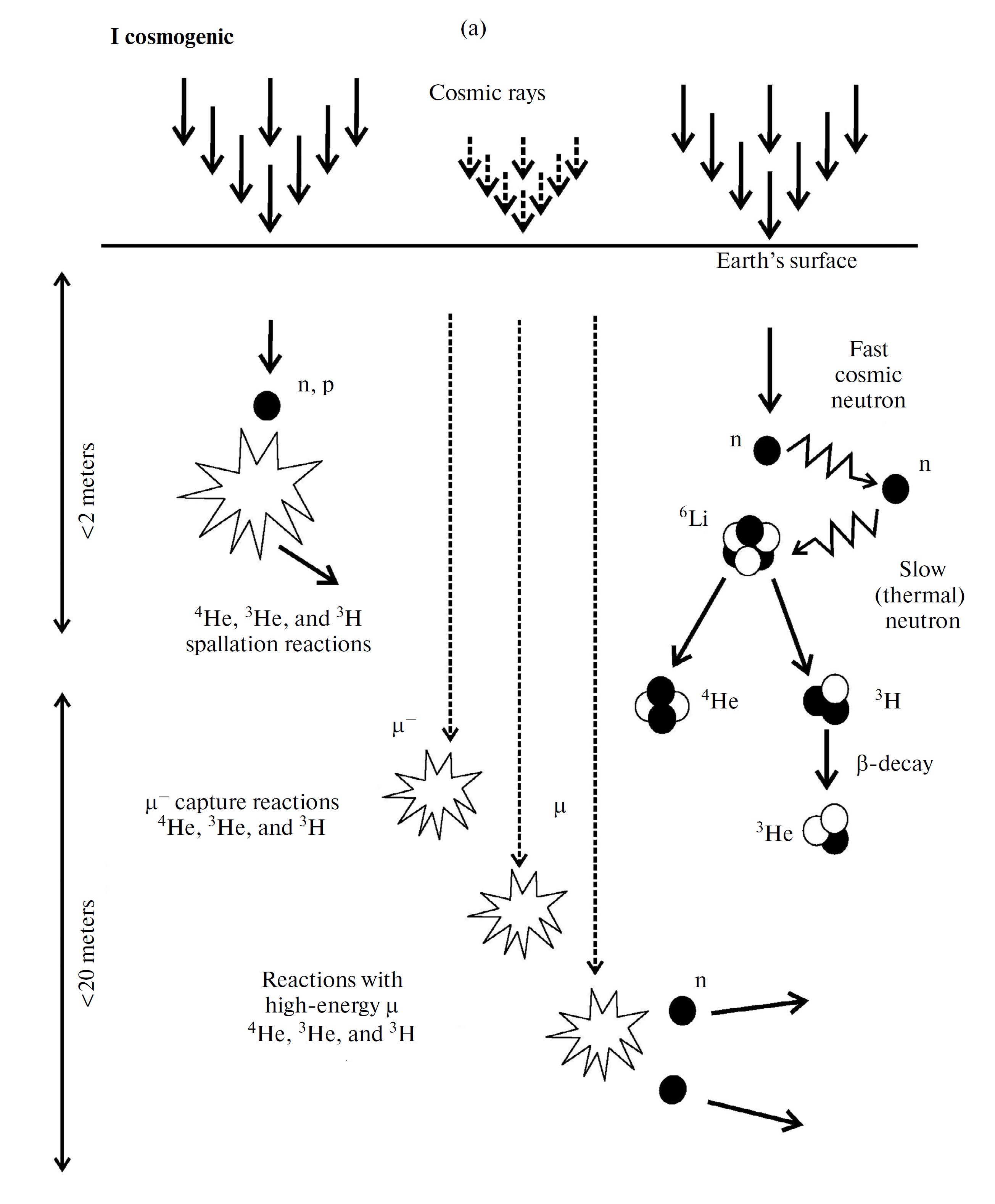}
		\caption{Schematic representation of $^3$He accumulation in a rock: (a) channels of $^3$He production in a near-surface layer, (b) channels of $^3$He production in the whole rock volume.}
		\label{fig1a}
\end{figure}

\begin{figure}[t]
	\ContinuedFloat
	\centering
	\includegraphics[width = 0.7\textwidth]{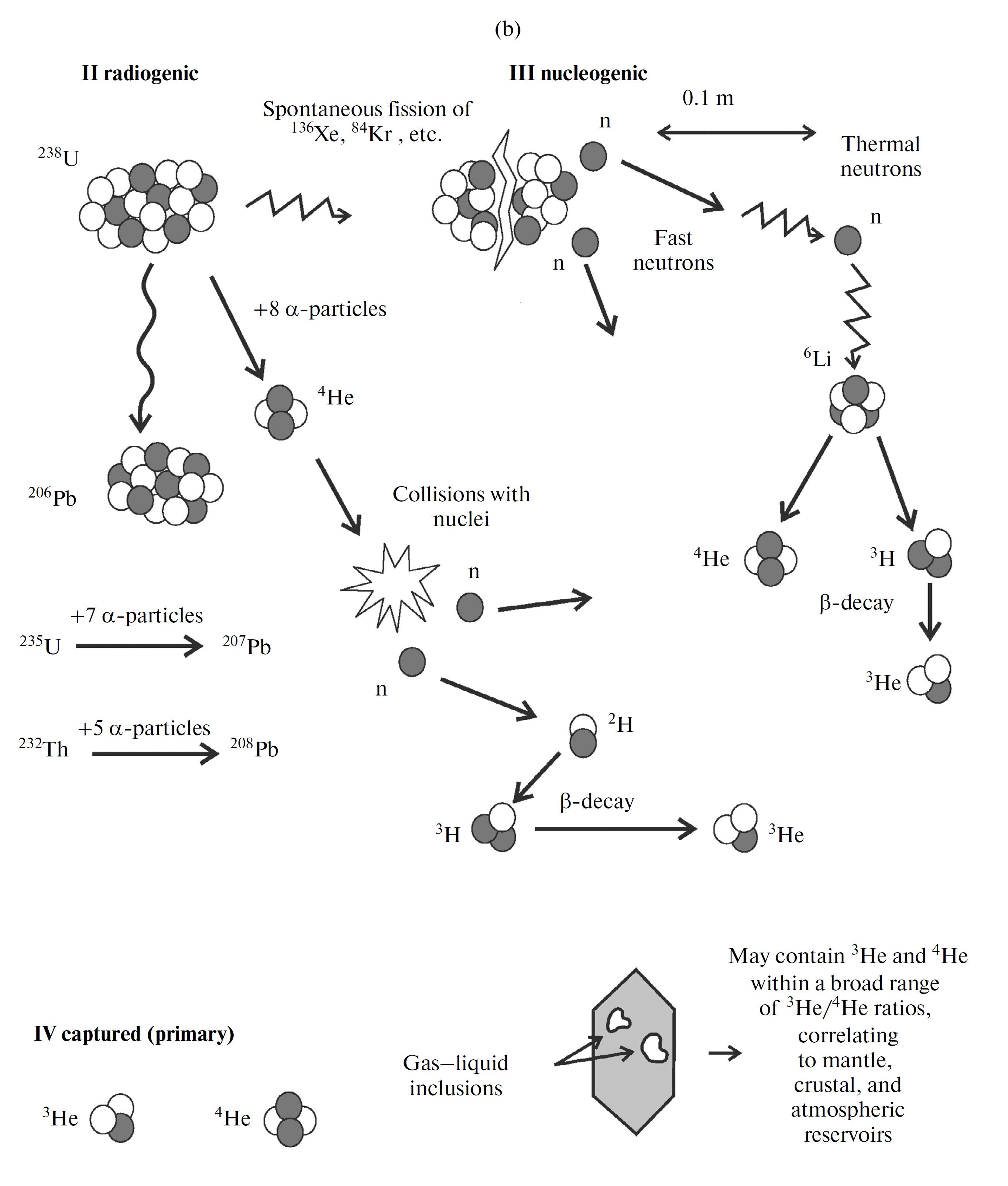}
		\caption{(Contd.)}
		\label{fig1b}
\end{figure}

\section{Cross Sections of Reactions Producing $^3$H and $^3$He}
\noindent
To calculate the production rates of cosmogenic nuclides, it is necessary to know the fluxes of cosmic ray particles and the cross sections of the nuclear reactions. Extensive information is now available on the cross sections of reactions in which nuclides are produced by protons in various targets, but such data are usually absent for neutron and muon induced reactions (Reedy, 2013; Heisinger et al., 2002a). We have calculated the cross sections of reactions and the energy distributions of $^3$H and $^3$He nuclei using the GEANT4 simulation toolkit. The data thus obtained were then utilized to evaluate the production rates of $^3$He in the ground by spallation reactions of cosmic-ray neutrons, reactions induced by high-energy muons, and $\mu^-$ capture reactions by atoms of the environment. In calculating the production rates of the nuclide by protons, we made use of available experimental data on the reaction cross sections.

\subsection{Numerical Simulation of the Nuclear Reactions}
The nuclear reactions were numerically simulated by the GEANT4, version 10.0, simulation toolkit (Agostinelli et al., 2003; Allison et al., 2006). Specialized commands were utilized in the numerical code to specify the model for the target and the energy of bombarding particles and physical processes of particle interaction. In our model, the target has a parallelepiped shape and is bombarded by high-energy particles: neutrons and muons. The chemical composition of the target material is assumed as the average composition of the upper crust (Rudnik and Gao, 2003) and is referred to as 'standard' ground hereafter. The contents of chemical elements taken into account in our simulations were as follows: 48.3 wt \% O, 31.1 wt \% Si, 8.2 wt \% Al, 3.6 wt \% Fe, 2.6 wt \% Ca, 2.4 wt \% Na, 2.3 wt \% K, and 1.5 wt \% Mg. The density of the ground was assumed to be 2.65 g/cm$^3$. The average
mass number and atomic number of the target material were 20.9 and 10.4, respectively. The production rates of the $^3$H and $^3$He nuclides in neutron-induced spallation reactions depend on the chemical composition of the ground: the larger the mass number of the target material, the lower the production rate (Masarik and Reedy, 1996; Farley et al., 2006). However, the calculation errors related to the inaccuracy of the nuclear interaction models are greater than the possible errors related to the composition of the ground, and hence, we have provisionally carried out our simulations only with the 'standard' ground.

Our model makes use of standard lists of physical processes and particle interaction models \verb|FTFP_BERT_HP| and \verb|QGSP_BIC_HP|. These physics lists make use of different intranuclear cascade models to describe the interaction of high-energy neutrons with nuclei: Bertini's intranuclear cascade model (\verb|FTFP_BERT_HP|) and the binary intranuclear cascade model (\verb|QGSP_BIC_HP|). The models differ in the descriptions of the intranuclear particle cascades and hadron-hadron interaction (Heikkinen et al., 2003; Folger et al., 2004). The interaction between muons and atoms in the \verb|FTFP_BERT_HP| and \verb|QGSP_BIC_HP| physics lists is described by similar models.

The cross sections of reactions producing nuclide $i$ induced by particles $k$ with energy $E$ can be calculated by the formula

\begin{equation}
	 \sigma_{ik} (E) = \frac{N_{ik}}{N_k C d},
	 \label{cs}
\end{equation}

\noindent
where $N_k$ is the number of projectile particles for which numerical simulation was conducted, $N_{ik}$ is the number of nuclei $i$ produced by nuclear reactions in the target, $C$ is the total concentration of atoms in the target material, and $d$ is the thickness of the target. If the material of the target contains more than one chemical element, then the cross section (\ref{cs}) is the average cross section of nuclide production in this material. In our numerical simulations, we have calculated the average cross sections of reactions for the 'standard' chemical composition of the ground.

The energy distribution of nuclei $i$ (reaction products) was calculated as

\begin{equation}
	p_{im}(E) = \frac{N_{im}}{N_i},
\end{equation}

\noindent
where $E$ is the energy of the projectile particle, $N_{im}$ is the number of nuclei $i$ produced by the reaction, with the energy lying within the range $[\varepsilon_m - \Delta \varepsilon_m, \varepsilon_m + \Delta \varepsilon_m]$ and $N_i$ is the total number of
produced $i$ nuclei.

In this model, the thickness of the target $d$ was assumed to be 1 cm for neutrons and 1 m for muons. The number of neutrons for which the numerical simulations were carried out was $3 \times 10^7$ for each particle
energy value. The value of this parameter for muons is $3 \times 10^6$. For high-energy particles, the condition of a thin target is valid: the free path length for neutrons before their interaction is much greater than the thickness of the target, and the ionization and radiative energy losses of the muon in the target are much lower than the muon energy.

\subsection{Cross Sections of Nuclide-Producing Reactions Induced by High-Energy Neutrons}
\begin{figure}[t]
	\centering
	\includegraphics[width = 0.65\textwidth]{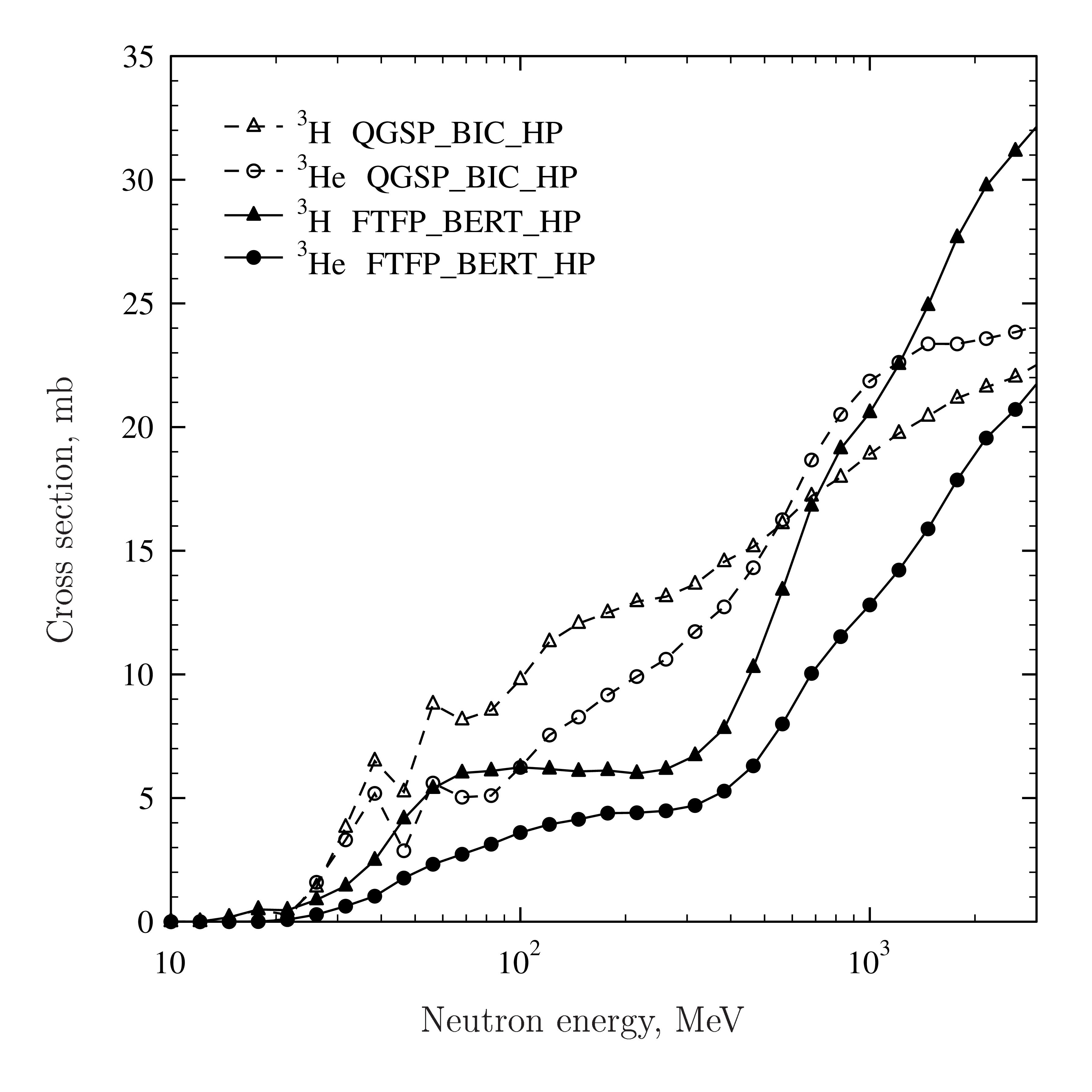}
		\caption{Cross sections of high-energy neutron-induced reactions producing $^3$H and $^3$He in the 'standard' ground. Values along the ordinate are in millibarns (mb), 1 b = $10^{-24}$ cm$^2$. The statistical calculation errors are approximately 1\% (1$\sigma$)}
		\label{fig2}
\end{figure}

\noindent
In spallation reactions, a high-energy nucleon collides with atomic nuclei of the material and knocks nucleons and light nuclei out of them. We have calculated the cross sections of reactions forming the nuclei $^3$H, $^3$He, and $^4$He induced by high-energy neutrons in the 'standard' ground. Figure \ref{fig2} displays the excitation functions for the reactions producing $^3$H and $^3$He. We have conducted these simulations for two physics lists: \verb|FTFP_BERT_HP| and \verb|QGSP_BIC_HP|. The kinetic energy of the produced nuclei lies within the broad range of 100 keV to 100 MeV.

The characteristic energy of high-energy cosmic-ray neutrons is approximately 100 MeV (Gordon et al., 2004). The ratio of the total cross section of reactions producing the $^3$H and $^3$He nuclides to the cross section of the reaction producing $^4$He is close to 0.1 for these neutron energy values. This value characterizes the isotopic ratio of cosmogenic He produced by spallation induced by cosmic-ray high-energy nucleons. Note that cosmogenic He in stony meteorites has a $^3$He/$^4$He ratio close to 0.2 (Wieler, 2002). In this situation, cosmogenic He is produced by nuclear reactions induced by protons, alpha particles of galactic cosmic rays, and by secondary neutrons and protons.

\subsection{Cross Sections of Nuclide Production Reactions by High-Energy Muons}

A high-energy muon spends its energy when entering a material to ionization of its atoms and molecules, bremsstrahlung, production of $e^+e^-$ pairs, and photo-nuclear interaction. These processes give rise to cascades of secondary particles. Nuclides can be produced by nuclear reactions induced by secondary particles.

\begin{figure}[t]
	\centering
	\includegraphics[width = 0.65\textwidth]{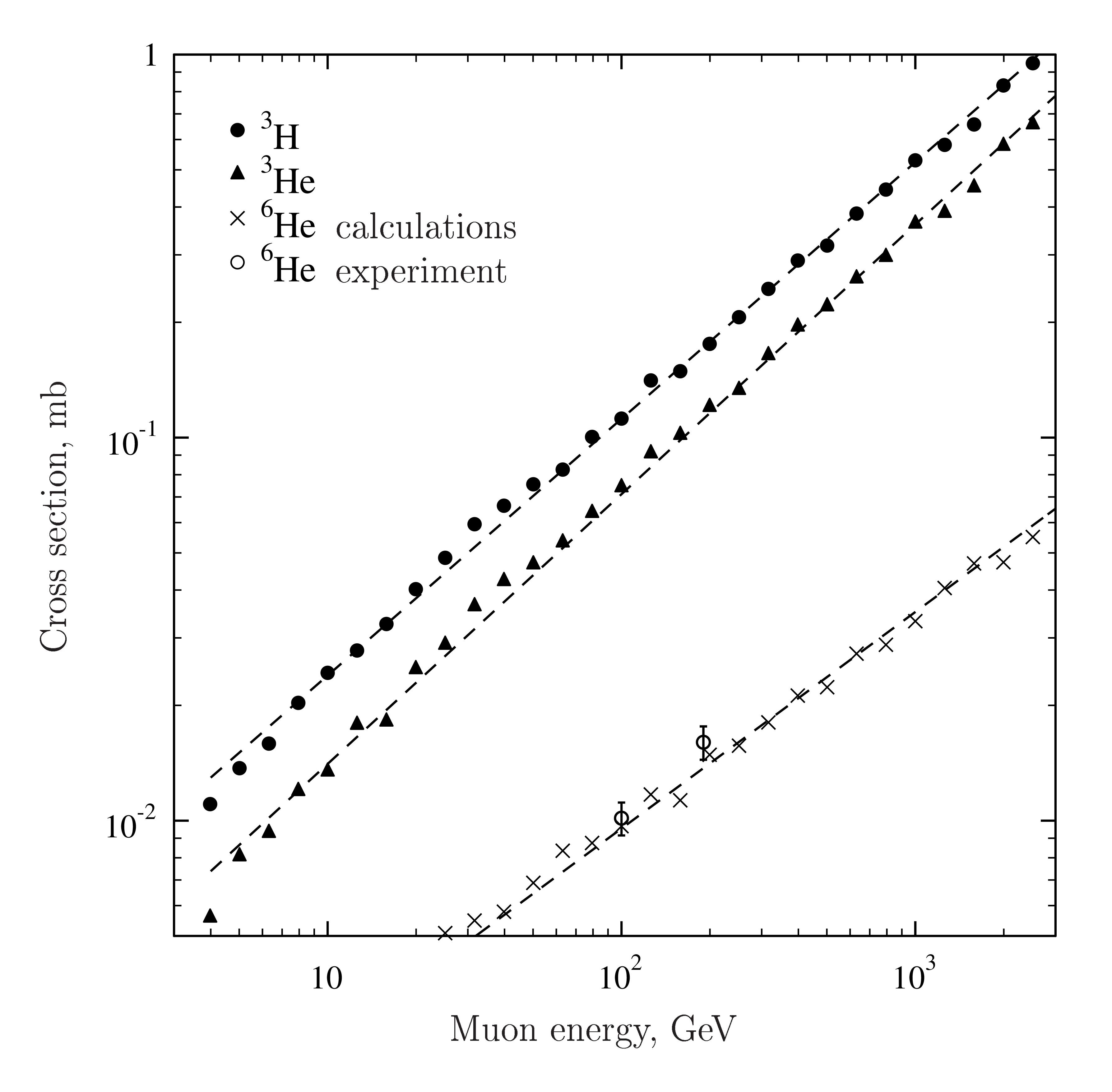}
		\caption{Cross sections of high-energy muon-induced reactions producing $^3$H and $^3$He in 'standard' ground. The dashed line is a power-law approximation. The figure also shows experimental data from (Heisinger et al., 2002a) and our calculation results for cross sections of $^6$He-producing reaction on carbon. The statistical calculation errors (1$\sigma$) are 2-7\% for $^3$H, 2-10\% for $^3$He, and 3-15\% for $^6$He.}
		\label{fig3}
\end{figure}

Figure \ref{fig3} shows the calculated cross sections of reactions producing the $^3$H and $^3$He nuclides in the 'standard' ground by high-energy muons. The simulations were made with the \verb|FTFP_BERT_HP| physics list. The excitation function can be approximated by a power function

\begin{equation}
	\sigma(E) = \sigma_0 E^\alpha,
	\label{csm}
\end{equation}
\noindent
where the value of the parameter $E$ corresponds to the energy of muons in GeV, and $\sigma_0$ is the cross section value at muon energy of 1 GeV. The calculated values of the $\sigma_0$ and $\alpha$ parameters are as follows: $\sigma_0$ = 0.0051 mb and $\alpha$ = 0.67 for the $^3$H nuclide, and $\sigma_0$ = 0.0028 mb and $\alpha$ = 0.70 for the $^3$He nuclide. According to our simulations, the ratio of the total cross section of reactions producing the $^3$H and $^3$He nuclides to the cross section of the reaction producing $^4$He is 0.05-0.1. If the \verb|QGSP_BIC_HP| physics list is employed in the simulations, the results are analogous, and the values of the cross sections are 10-20\% greater.

The production of cosmogenic nuclides by high-energy muons in various materials was experimentally studied in (Heisinger et al., 2002a). The cross sections of the reactions were measured for two energy values of muons: 100 and 190 GeV. The measured cross section values for various materials and nuclides broadly vary from 0.001 to 10 mb, whereas the parameter $\alpha$ lies within the range of 0.5-1.5. The measured cross sections of the reaction in which $^6$He is produced by high-energy muons on carbon are reported in (Heisinger et al., 2002a). To test the numerical model, we have calculated the cross sections of this reaction (Figure \ref{fig3}) using the \verb|FTFP_BERT_HP| physics list. The calculated and experimental data are in reasonably good agreement.

\subsection{Cross Sections of $\mu^-$ Capture Reactions}
When passing through a material, a muon is eventually captured by one of the atoms of this material. The muon passes down to the $1s$ energy level, and this process is associated with the emission of X-ray photons and Auger electrons. After this, the muon either decays to an electron and neutrino or interacts with the nucleus (Heisinger et al., 2002b). In nuclear reactions, the newly produced excited nucleus passes to its ground state and emits gamma-quanta, nucleons, and light nuclei.

We have simulated $\mu^-$ capture in the 'standard' ground. Let $f_i$ be the fraction of muons that stop in the ground and induce nuclear reactions that produce nuclide $i$. The values of $f_i$ calculated for the $^3$H, $^3$He, and $^4$He nuclides are 0.0041, 0.00036, and 0.055, respectively. The statistical errors of the calculations (1$\sigma$) are 1, 3, and 0.3\%, respectively. The calculated average kinetic energy values of the produced $^3$H, $^3$He, and $^4$He nuclei are 7, 10, and 5 MeV, respectively.

According to our evaluations, the average number of neutrons produced in $\mu^-$ capture is $f_n$ = 0.75. The average energy of such neutrons is close to 4 MeV.

\section{Production Rate of Cosmogenic $^3$He in Ground}

\subsection{Production Rate of $^3$He by High-Energy Nucleons}

Below we calculate the production rates of the $^3$H and $^3$He nuclides in a near-surface ground layer by high-energy cosmic-ray neutrons and protons. The production rate of nuclide $i$ in reactions induced by particles $k$ is

\begin{equation}
	P_{i k} = C \int \limits_0^\infty dE \sigma_{ik}(E) I_k(E)
\end{equation}

\noindent
where $I_k(E)$ is the angle-integrated differential flux of particles $k$, $\sigma_{ik}(E)$ is cross section of a reaction producing the nuclide, and $C$ is the volume concentration of atoms.

In calculating the production rates of the $^3$H and $^3$He nuclides in spallation reactions by cosmic-ray neutrons, we utilized the cross sections of reactions evaluated by means of numerical simulation. The angle-integrated differential neutron flux in the Earth's atmosphere near the surface was compiled from (Gordon et al., 2004). This energy spectrum of neutrons is an approximation of measurements and pertains to high geomagnetic latitudes, sea level, and an average level of solar activity. The error of the neutron energy spectrum is no higher than 10-15\% (Gordon et al., 2004). The table reports the calculated production rates of the $^3$H and $^3$He nuclides induced by cosmic-ray neutrons in the near-surface layer of the ground. The ratio of the $^3$H and $^3$He production rates is approximately 1.5.

~\\
~\\
\begin{center}
\begin{tabular}{l | c | c}
\multicolumn{3}{l}{Table. Production rates of nuclides in spallation reactions induced by cosmic-ray neutrons} \\ [5pt] 
\hline \\ [-2ex]
Reaction cross section & $^3$H/$^3$He production rate ratio & \parbox[t]{5.5cm}{Production rate $^3$H + $^3$He atoms g$^{-1}$ year$^{-1}$} \\ [20pt] 
\hline \\ [-2ex]
\verb|FTFP_BERT_HP| & 1.7 & 35 \\ [3pt]
\verb|QGSP_BIC_HP| & 1.3 & 60 \\ [3pt]
\hline
\end{tabular}
\end{center}
~\\
~\\

In calculating the production rate of the $^3$He nuclide by cosmic-ray protons, we utilized experimental data on the cross sections of reactions on Si, Al, Mg (Michael et al., 1989, 1995; Leya et al., 1998; Demetriou et al., 2005), O (Bertrand and Peelle, 1973), and Fe (Ammon et al., 2008). The production of $^3$He by protons is dominated by reactions on Si and O. For the reaction O(p,X)$^3$He, one experimental value of the cross section was reported in (Bertrand and Peelle, 1973), and hence, the calculated production rate of $^3$He by protons should be regarded as an approximate estimate. The angle-integrated differential proton fluxes were borrowed from (Sato et al., 2008; Nesterenok, 2013). The energy spectra of protons were obtained in these publications by means of numerical simulation of cosmic ray propagation through the Earth's atmosphere. The energy spectra in (Sato et al., 2008) and (Nesterenok, 2013) are practically identical and lead to closely similar production rates of the nuclide. The ratio of the production rates of the $^3$H and $^3$He nuclides was assumed in the simulations to be 1 (Leya et al., 2004). According to our evaluations, the total production rate of $^3$H and $^3$He by cosmic-ray protons in a near-surface ground layer at an average solar activity level at high geomagnetic latitudes and at sea level is approximately 10 atoms g$^{-1}$ year$^{-1}$.

The calculated production rates of $^3$He (including its precursor $^3$H) by cosmic-ray neutrons and protons are approximately 45 and 70 atoms g$^{-1}$ year$^{-1}$ in a near-surface ground layer at sea level and high geomagnetic latitudes (the two values correspond to the physics lists used to calculate the cross sections of neutron-induced spallation reactions). For comparison, the average $^3$He production rate in olivine and pyroxene obtained by measured concentrations of the nuclide in rocks is about 130 atoms g$^{-1}$ year$^{-1}$ at sea level and high geomagnetic latitudes (Goehring et al., 2010). Numerical simulations of nuclear reactions still cannot be used to accurately calculate the cross sections of nuclide-producing reactions, but results of such numerical simulations can be employed to calculate the ratio of the production rates of the nuclides $^3$H/$^3$He and the energy distribution of the nuclei.

\begin{figure}[t]
	\centering
	\includegraphics[width = 0.65\textwidth]{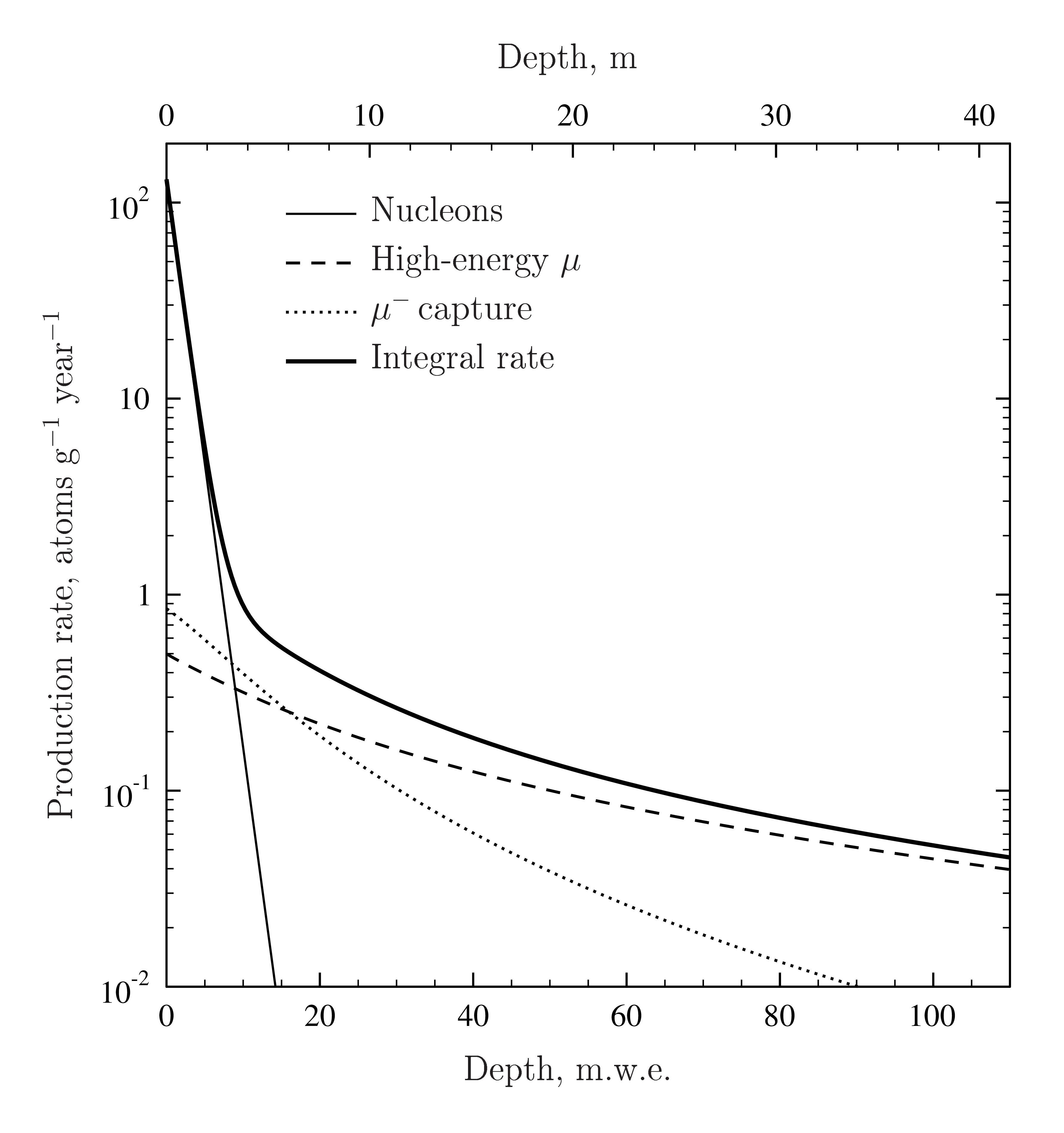}
		\caption{Production rates of cosmogenic $^3$He (including precursor $^3$H) in 'standard' ground. Values along the abscissa show: bottom -- depth in m.w.e. (1 m.w.e. = 100 g/cm$^2$), top -- depth in m for a ground 2.65 g/cm$^3$ in density. The calculation results correspond to an average solar activity level, high geomagnetic latitudes, and sea level.}
		\label{fig4}
\end{figure}

For the production rate of a nuclide induced by high-energy cosmic-ray nucleons at depth $z$ in the ground, it can be written (Masarik and Reedy, 1995)

\begin{equation}
	P_{i, h}(z) = P_{i, h}^{0} \mathrm{exp}(-z/\Lambda_h)
\end{equation}

\noindent
where $z$ is the depth of the ground layer in mass depth units; $P_{i, h}^{0}$ is the production rate of nuclide $i$ in the near-surface layer, $\Lambda_h$ is the thickness of the ground at which the production rate of the nuclide decreases by a factor of $e \approx 2.72$. Mass depth is $z = \rho h$ where $\rho$ is the average density of the ground, and $h$ is the depth in cm. The unit of measure of mass depth is g/cm$^2$ or meters of water equivalent (m.w.e.), 1 m.w.e. = 100 g/cm$^2$. According to measured concentrations of cosmogenic nuclides in rock samples, $ \Lambda_h = 1.5-1.8$ m.w.e. (Farley et al., 2006; Dunai, 2010). Figure \ref{fig4} shows the production rate of the $^3$He nuclide (including its precursor $^3$H) induced by high-energy nucleons in the ground as a function of depth $z$. We assumed the following values of the parameters: $P_{i, h}^{0}$ = 130 atoms g$^{-1}$ year$^{-1}$ and $\Lambda_h$ = 1.5. m.w.e.

The flux of cosmic-ray neutrons and protons near the Earth's surface strongly depends on the thickness of the atmosphere and geomagnetic coordinates. A number of models were suggested to describe the dependence of the production rate of the nuclide in the ground on the altitude of the location above sea level and its geomagnetic coordinates (Desilets and Zreda, 2003; Dunai, 2010; Lifton et al., 2014).

\subsection{Production Rate of $^3$He by High-Energy Muons}
The production rate of nuclide $i$ by high-energy muons at depth $z$ in the ground can be approximately expressed as (Heisinger et al., 2002a)

\begin{equation}
	P_{i, {\mu}fast}(z) = \sigma_0 C I_{\mu}(z) \beta(z) \langle E_{\mu} \rangle^{\alpha}
	\label{eqn_mu_fast}
\end{equation}

\noindent
where $\sigma_0$ and $\alpha$ are the cross section and exponent for nuclide $i$ according to (\ref{csm}), $C$ is the volume concentration of atoms in the ground, $I_{\mu}(z)$ is the angle-integrated differential muon flux at depth $z$, $\langle E_{\mu} \rangle$ is the average energy of muons at depth $z$ in GeV, the function $\beta(z)$ weakly depends on depth, $\beta(z) \approx 0.85$. Expressions for $I_{\mu}(z)$, $\langle E_{\mu} \rangle$, and $\beta(z)$ are given in (Heisinger et al., 2002a).

Figure \ref{fig4} shows the production rate of the nuclide $^3$He (including its precursor $^3$H) by high-energy muons in the ground at sea level. The depth $z$ at which the production rate of $^3$He decreases by a factor of $e \approx 2.72$ is 25 m.w.e. (or 9.5 m for a ground whose density is 2.65 g/cm$^3$).

The calculation of the production rate of nuclides in a ground by muons at different altitudes and geomagnetic latitudes is discussed in (Desilets and Zreda, 2003; Nesterenok and Naidenov, 2012a; Lifton et al., 2014).

\subsection{Production of $^3$He in $\mu^-$ Capture Reactions}

The production rate of nuclide $i$ in $\mu^-$ capture reactions at depth $z$ in the ground is

\begin{equation}
	P_{i, \mu^-}(z) = f_i f_{\mu^-} \Big\vert \frac{dI_{\mu}(z)}{dz} \Big\vert,
\end{equation}

\noindent
where $ f_{\mu^-}$ = 0.44 is the fraction of negative muons in the muon flux (Heisinger et al., 2002b). Figure \ref{fig4} displays the production rate of the nuclide $^3$He (including $^3$H) depending on depth. The depth at which the production rate decreases by a factor of $e \approx 2.72$ is 13 m.w.e.

\subsection{Energy Distribution of $^3$H and $^3$He Nuclei}

\begin{figure}[t]
	\centering
	\includegraphics[width = 0.65\textwidth]{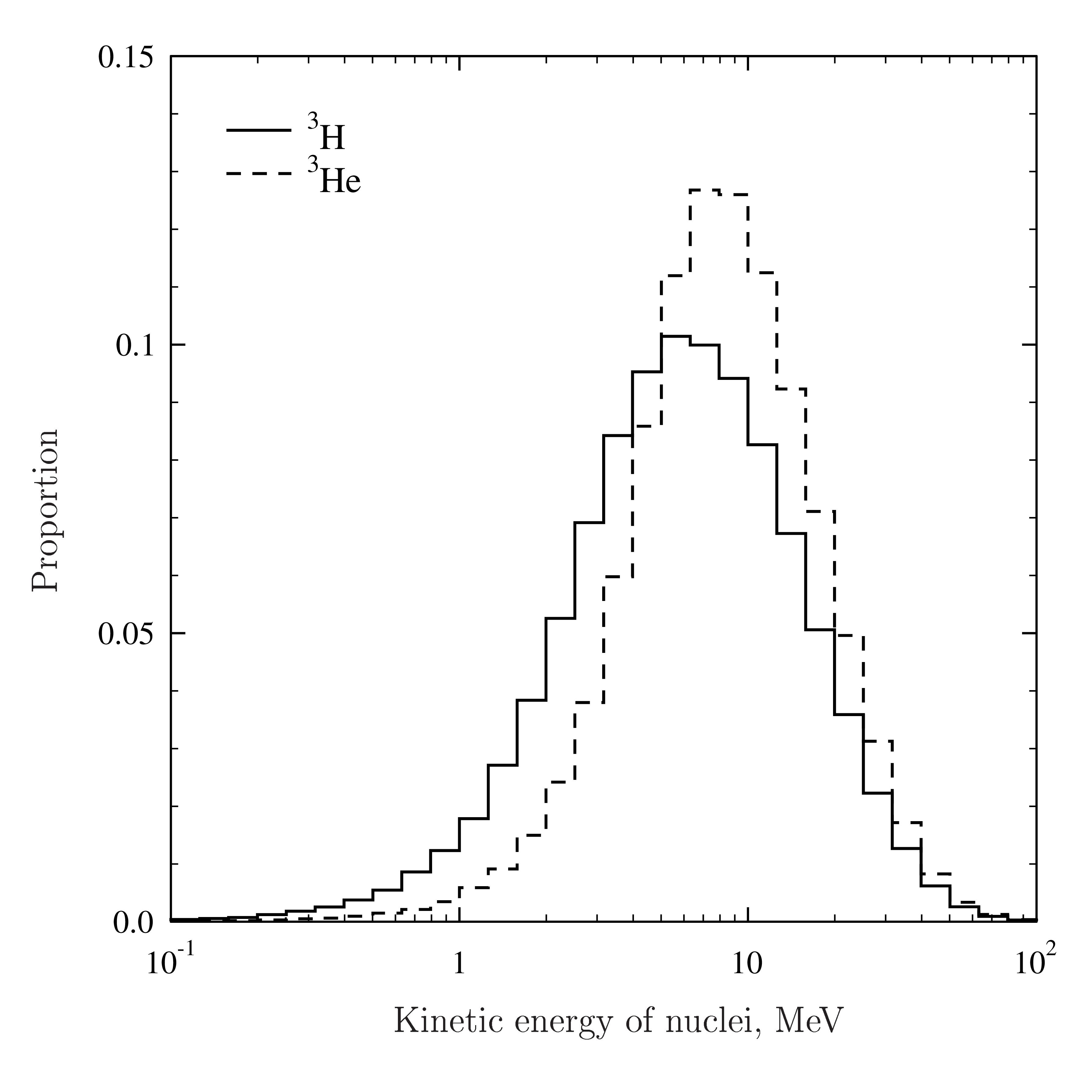}
		\caption{Energy distribution of nuclei produced in spallation reactions induced by cosmic-ray neutrons. Values along the ordinate correspond to the proportion of nuclei whose energy belongs to the energy interval. The number of energy intervals per one order of magnitude is ten; the intervals are equal in length on logarithmic scale. The statistical errors of the calculations are 1\% (1$\sigma$).}
		\label{fig5}
\end{figure}

The energy distribution of $^3$H and $^3$He nuclei produced in spallation reactions induced by cosmic-ray
neutrons can be calculated as

\begin{equation}
\overline{p}_{im} = \frac{1}{P_{i,n}} C \int\limits_0^{\infty} dE \, p_{im}(E) \sigma_{in}(E) I_n(E) ,
\end{equation}

\noindent
where $\overline{p}_{im}$ is the fraction of nuclei $i$ ($^3$H or $^3$He) whose energy is in the energy range $m$, $E$ is the energy of neutrons, $I_n(E)$ is the angle-integrated differential neutron flux, and $\sigma_{in}(E)$ is the cross section of the reaction producing the nuclide, the function $p_{im}(E)$ was calculated in numerical simulations [see (2)]; the distribution is normalized to the production rate of the nuclide by neutrons $P_{i,n}$. Figure \ref{fig5} displays the calculated energy distributions of $^3$H and $^3$He nuclei. The calculation results are presented for the \verb|FTFP_BERT_HP| physics list. The average nucleus energy is approximately 10 MeV for both nuclides.

\section{$^3$H Production in Reaction on Li}

The source of $^3$He in rocks is thermal neutron capture reaction $^6$Li(n,$\alpha$)$^3$H with subsequent $^3$H $\beta$-decay into $^3$He (Figure \ref{fig1a}).

Neutrons are produced in the ground by nuclear reactions involving alpha-particles, which are produced as a result of U and Th radioactive decays (Feige et al., 1968). For the relative concentrations of chemical elements used in our simulations, the production rate of neutrons is approximately 1.3 year$^{-1}$ g$^{-1}$ per 1 $\mu$g g$^{-1}$ of U and 0.6 year$^{-1}$ g$^{-1}$ per 1 $\mu$g g$^{-1}$ of Th (Feige et al., 1968; Andrews, 1985). Neutrons can also be formed by spontaneous fission reactions of the $^{238}$U isotope. The production rate of neutrons in this channel is 0.5 year$^{-1}$ g$^{-1}$ per 1 $\mu$g g$^{-1}$ U (Feige et al., 1968; Shukolyukov, 1970). Assuming the U content to be 3 $\mu$g g$^{-1}$ and that of Th to be 11 $\mu$g g$^{-1}$ (Rudnik and Gao, 2003), for the production rate of neutrons we obtain $P_{n,UTh} \approx$ 10 year$^{-1}$ g$^{-1}$. Note that the production rate of neutrons in rocks where U and Th are not uniformly distributed can be much lower than these estimates (Martel et al., 1990). The nuclide produced in reactions involving radiogenic neutrons is referred to as nucleogenic.

Neutrons are produced in nuclear reactions induced by cosmic-ray high-energy muons. According to (Heisinger et al., 2002a), the production rate of neutrons by high-energy muons can be written as

\begin{equation}
	P_{n, {\mu}fast}(z) = 4.8 \times 10^{-6} \mathrm{g} ^{-1} \mathrm{cm}^2 \times \beta(z) I_{\mu}(z) \langle E_{\mu} \rangle^{\alpha}
\end{equation}

\noindent
The production rate of neutrons in $\mu^-$ capture reactions is

\begin{equation}
	P_{n, \mu^-}(z) = f_n f_{\mu^-} \Big\vert \frac{dI_\mu(z)}{dz} \Big\vert
\end{equation}

In near-surface ground layers, the source of thermal neutrons is the nuclear active component of cosmic rays. As a rough approximation, the production rate of thermal neutrons of cosmic rays can be written as (Phillips et al., 2001)

\begin{equation}
	P_{n,h}(z) = P^0_{n,h} \mathrm{exp}(-z/\Lambda_h).
\end{equation}

\noindent
The production rate of epithermal neutrons (whose energy is lower than 1 keV) in a near-surface ground layer at an average level of solar activity at high geomagnetic latitudes and at sea level is $P^0_{n,h} \approx$ 750 year$^{-1}$ g$^{-1}$ (Phillips et al., 2001). The absorption of epithermal neutrons by material atoms and the diffusion of thermal neutrons from the ground to atmosphere were ignored in our evaluations of the production rate of the nuclide.

For the production rate of $^3$H in the reaction $^6$Li(n,$\alpha$)$^3$H, it can be written

\begin{equation}
	P_{^3H}(z) = k F_\mathrm{Li}P_n(z)
\end{equation}

\noindent
where $P_n(z)$ is the production rate of neutrons in various processes, $k$ is the fraction of neutrons relative to the total number of newly produced neutrons in the environment that reach thermal energies, $k \approx 0.8-1$ (Morrison and Pine, 1955; Ballentine and Burnard, 2002), and $F_\mathrm{Li}$ is the fraction of thermal neutrons that react with Li. The parameter $F_\mathrm{Li}$ depends on the relative concentrations of Li and chemical elements possessing large cross sections of thermal neutron capture (B, Gd, Sm, Cl, etc.) in the rock. According to data in (Andrews and Kay, 1982), $F_\mathrm{Li} \approx 0.001 \times$ [Li], where [Li] is the Li concentration in ppm. The $^3$H production rate is proportional to the Li concentration in the rock that broadly varies for various rocks. The average Li concentration in the upper crust is approximately 35 ppm (Teng et al., 2004), and we made use of this value in our simulations.

\begin{figure}[t]
	\centering
	\includegraphics[width = 0.65\textwidth]{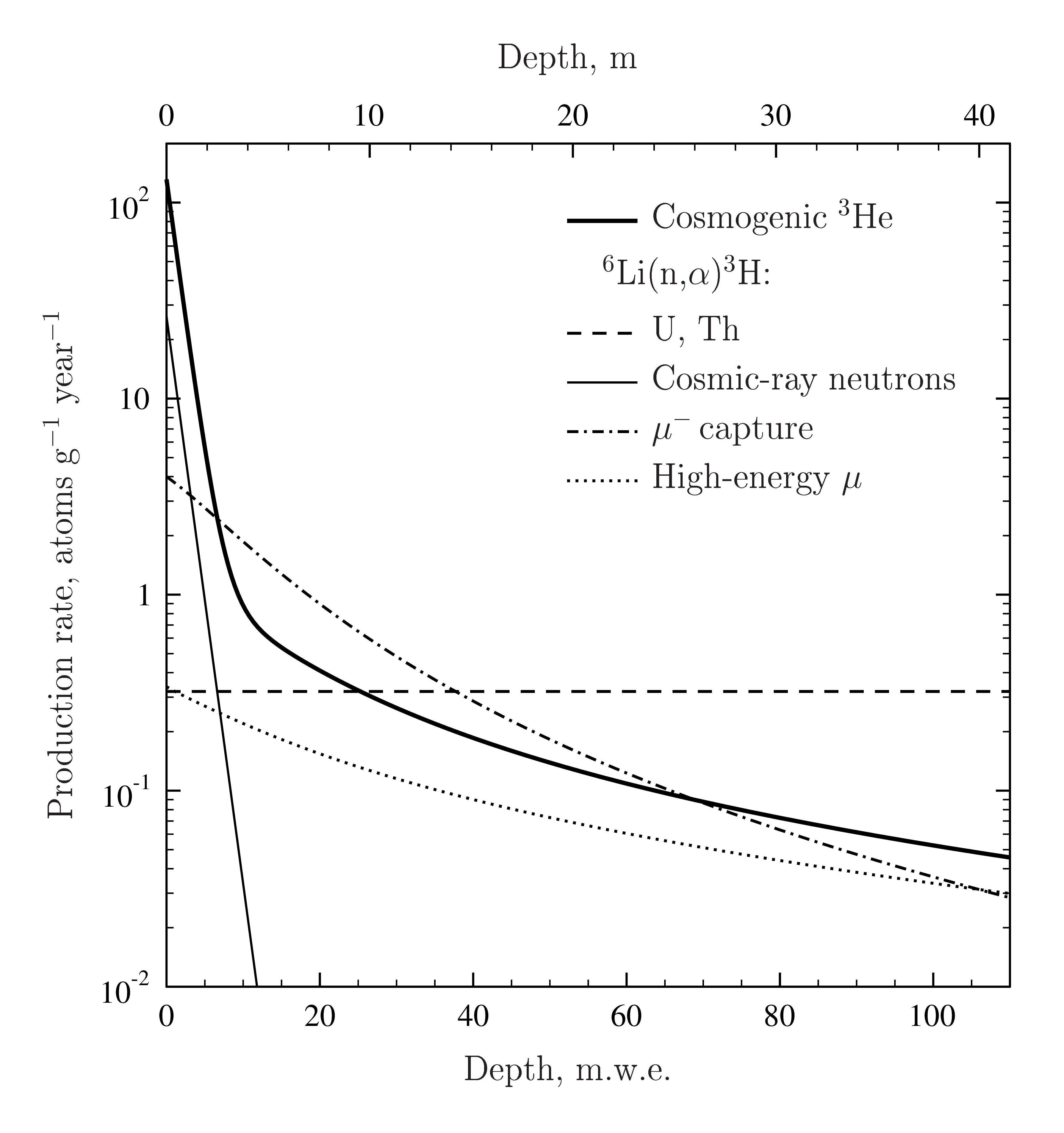}
		\caption{Production rate of $^3$H in the reaction $^6$Li(n,$\alpha$)$^3$H. For comparison, the figure also shows the production rate of cosmogenic $^3$He. The calculation results pertain to an average level of solar activity, high geomagnetic latitudes, and sea level.}
		\label{fig6}
\end{figure}

Figure \ref{fig6} exhibits the calculated production rate of $^3$H in the reaction $^6$Li(n,$\alpha$)$^3$H. Cosmic-ray thermal neutrons produce $^3$H in the upper ground layer no thicker than 10 m.w.e. At depths of 10-40 m.w.e., $^3$H is generated mostly under the effect of neutrons resulting from negative muon capture. At even greater depths, the nuclide is produced first of all by neutrons produced during U and Th decays.

\section{Discussion}

Spallation reactions induced by cosmic-ray high-energy neutrons and protons are the dominant channel of $^3$He production in the near-surface ground layer. The estimated $^3$He production rate in spallation reactions induced by cosmic-ray protons accounts for approximately 10\% of the total production rate of cosmogenic $^3$He at sea level and high geomagnetic latitudes. Proton and neutron fluxes in the Earth's atmosphere differently depend on altitude and geomagnetic latitude. The contribution of protons to the overall cosmic-ray nucleon flux increases with increasing altitude above sea level and decreasing geomagnetic latitude (Sato et al., 2008; Nesterenok, 2013). According to our evaluations, the contribution of protons to the total $^3$He production rate may amount to 15-20\% at altitudes of 4-5 km and low geomagnetic latitudes (Himalayas, northern and central Andes). Thus, the dependence of the $^3$He production rate on the altitude of a locality and its geomagnetic coordinates is different from the analogous dependences of other cosmogenic nuclides, such as in-situ $^{14}$C, for which the proton-induced production channel is insignificant (Nesterenok and Naidenov, 2012b). This should be taken into account in 'calibrating' the $^3$He production rate in rocks by comparing data on concentrations of $^3$He and other cosmogenic nuclides. For example, anomalously high $^3$He/$^{10}$Be ratios in rock samples from the Himalayas were reported in (Gayer et al., 2004; Amidon et al., 2008). Current scaling models for production rates of cosmogenic nuclides at various altitudes above sea level and geomagnetic coordinates require certain adjustments when applied to $^3$He (Lifton et al., 2014).

The contribution of muon-induced reactions to the overall production rate of cosmogenic $^3$He is approximately 1\% in the near-surface ground layer at sea level and at high geomagnetic latitudes. However, at depths greater than 10 m.w.e., the muon component is dominant in the production rate of the $^3$He cosmogenic nuclide. In an arid cold climate, the erosion rate is approximately 5-10 cm/Ma (Margerison et al., 2005). In this situation, if the rocks are exposed at the surface for a long enough time (10-100 Ma), much $^3$He in near-surface ground layers is produced by cosmic-ray muons. It is also expedient to consider the contribution of the muogenic component in studying rocks that are overlain from time to time by a relatively thin snow cover. In this case, this channel producing cosmogenic He in rocks is not shielded, in contrast to channels producing $^3$He in reactions induced by cosmic-ray neutrons and protons.

The production rate of $^3$H in the near-surface ground layer in thermal neutron capture reaction with Li is approximately 20\% of the production rate of the cosmogenic $^3$He nuclide at the Li concentration assumed in our simulations. Depending on the Li concentration in the rocks, this proportion may vary from 5 to 50\% (Dunai et al., 2007). At depths $z > $ 10 m.w.e., the production rate of $^3$H in reaction on Li is dominant in the overall production rate of the $^3$He nuclide for the ground model discussed herein.

If the production rates of the nuclide are different in two minerals, nuclei produced in one of the minerals can be ejected into grains of the other, and hence, result in a redistribution of the reaction products between the minerals (Farley et al., 2006). In this situation, the measured production rates should be different from the true ones. To evaluate the extent of implantation or ejection from one mineral to the other, one should know both the ratio of the production rates of $^3$H and $^3$He and their energy distributions. According to our evaluations, the average kinetic energy of $^3$H and $^3$He nuclei produced in neutron-induced spallation reactions and muon-induced reactions is about 10 MeV. For a $^3$He nucleus possessing kinetic energy of 10 MeV, the path length in the ground is 70 $\mu$m, and this length for a $^3$H nucleus of the same energy is 270 $\mu$m (Ziegler et al., 2010). The path lengths were calculated using the parameters of the ground assumed in our study. These lengths are commensurable with the grain sizes used in the studies. It is also worth mentioning that the behaviour of nuclides in their tracks is different: $^3$H can form chemical bonds, whereas $^3$He can be lost from a mineral due to diffusion within the track (Tolstikhin et al., 1999). Our estimates for the production rates of the nuclei and their energy distributions make it possible to more accurately describe the processes of nucleus ejection and implantation in minerals.

In our opinion, a promising approach to studying $^3$He in magmatic and metamorphic complexes is measuring the concentration profiles of the nuclide at depths up to 20-40 m from the surface. It is thereby important to utilize minerals that can relatively well preserve He. In cosmogenic-He dating of minerals, these minerals should not necessarily contain radioactive isotopes (U, Th, Sm, and others), and hence, along with minerals well-known in He thermochronology, such as zircon and apatite, these techniques can also be applied to such relatively 'geochronologically new' minerals as magnetite, hematite, and titanite, which are known to well preserve He (Starik, 1961; Lippolt et al., 1993). Considered together with theoretical calculations, an experimentally measured depth profile of $^3$He concentration provides information on the relative contents of magmatic, nucleogenic, cosmogenic, and muogenic components in the overall $^3$He budget in the rock and thus makes it possible not only to date the surface exposure of the rock but also to reproduce the erosion record in the area. Coupled application of this approach and conventional thermochronologic techniques (U-Th-He system of apatite and zircon and fission track dating) enables one to calculate the thermal parameters and date the surface exposure of the rock complexes. In practice, an indisputable advantage of this approach is the relative simplicity of analysis of minerals for $^3$He, which does not require accelerator mass spectrometry, in contrast to analysis for $^{36}$Cl, $^{10}$Be, and other cosmogenic radionuclides.

\section{Conclusions}

(1) Cross sections are calculated for neutron- and muon-induced reactions producing the $^3$He and $^3$H nuclides. Data presented in this publication make it possible to calculate the accumulation rate of $^3$He in a rock depending on its depth.

(2) Pioneering data are obtained that demonstrate that $^3$He production in spallation reactions induced by cosmic-ray protons makes up approximately 10\% of the overall production rate of cosmogenic $^3$He in the near-surface ground layer at sea level and high geomagnetic latitudes, which should be taken into account in scaling the production rate of the cosmogenic nuclide $^3$He at various areas.

(3) It is demonstrated that at depths of about 10-50 m.w.e., the muogenic component can significantly contribute the $^3$He accumulation. The production of $^3$He in this channel should be taken into consideration in interpreting experimental data on rocks that have been exposed at the surface for 10-100 Ma.

(4) An energy distribution is calculated for $^3$He and $^3$H nuclei produced in spallation reactions induced by cosmic-ray neutrons. These results can be utilized to calculate the extent of implantation and ejection of nuclei from one mineral to another.

(5) An approach is suggested for quantifying the exposure times of metamorphic and magmatic rock complexes on the Earth's surface on the basis of depth profiles of $^3$He concentration.

\section*{Acknoledgements}

This study was financially supported by the President's Council of Advisers on Research Grants (Grant MK-4760.2015.5), Program of Fundamental Research under the Earth's Science Division of the Russian Academy of Sciences 'Isotopic Systems in Geochemistry and Cosmochemistry. Methodological and Theoretical Aspects. Application in Evaluating the Parameters and Chronology of Geological Processes, Including Those in the Early Earth', and Program of the President of the Russian Federation for Support of Science Schools (Grant NSH-294.2014.2).

\section{References}

\textit{Agostinelli, S., Allision, J., Amako, K., et al.}, Geant4 -- a simulation toolkit, Nucl. Instrum. Methods Phys. Res. A, 2003, vol. 506, pp. 250-303.

\noindent
\textit{Allison, J., Amako, K., Apostolakis, J., et al.}, Geant4 developments and applications, IEEE Trans. Nucl. Sci., 2006, vol. 53, no. 1, pp. 270-278.

\noindent
\textit{Amidon, W.H., Farley, K.A., Burbank, D.W., and Pratt-Sitaula, B.}, Anomalous cosmogenic $^3$He production and elevation scaling in the High Himalaya, Earth Planet. Sci. Lett., 2008, vol. 265, pp. 287-301.

\noindent
\textit{Amidon, W.H., Rood, D.H., and Farley, K.A.}, Cosmogenic $^3$He and $^{21}$Ne production rates calibrated against $^{10}$Be in minerals from the Coso volcanic field, Earth Planet. Sci. Lett., 2009, vol. 280, pp. 194-204.

\noindent
\textit{Ammon, K., Leya, I., Lavielle, B., et al.}, Cross sections for the production of helium, neon and argon isotopes by proton-induced reactions on iron and nickel, Nucl. Instrum. Methods Phys. Res., Sect. B, 2008, vol. 266, no. 1, pp. 2-12.

\noindent
\textit{Andrews, J.N. and Kay, R.L.F.}, Natural production of tritium in permeable rocks, Nature, 1982, vol. 298, pp. 361-363.

\noindent
\textit{Andrews, J.N.}, The isotopic composition of radiogenic helium and its use to study groundwater movement in confined aquifers, Chem. Geol., 1985, vol. 49, pp. 339-351.

\noindent
\textit{Ballentine, C.J. and Burnard, P.G.}, Production, release and transport of noble gases in the continental crust, Rev. Mineral. Geochem., 2002, vol. 47, pp. 481-538.

\noindent
\textit{Bertrand, F.E. and Peelle, R.W.}, Complete hydrogen and helium particle spectra from 30- to 60-MeV proton bombardment of nuclei with A = 12 to 209 and comparison with intranuclear cascade model, Phys. Rev., 1973, vol. 8, pp. 1045-1064.

\noindent
\textit{Bierman, P.R.}, Using in situ produced cosmogenic isotopes to estimate rates of landscape evolution: a review from the geomorphic perspective, J. Geophys. Res., 1994, vol. 99, no. B7, pp. 13885-13896.

\noindent
\textit{Blard, P.-H., Pik, R., Lav\'{e}, J., et al.}, Cosmogenic $^3$He production rates revisited from evidences of grain size dependent release of matrix-sited helium, Earth Planet. Sci. Lett., 2006, vol. 247, pp. 222-234.

\noindent
\textit{Blard, P.-H. and Farley, K.A.}, The influence of radiogenic $^4$He on cosmogenic $^3$He determinations in volcanic olivine and pyroxene, Earth Planet. Sci. Lett., 2008, vol. 276, pp. 20-29.

\noindent
\textit{Demetriou, P., Dufauquez, Ch., Masri, Y.El., and Koning, A.J.}, Light charged-particle production from proton and alpha induced reactions on Si-natural at energies from 25 to 65 MeV: a theoretical analysis, Phys. Rev., 2005, vol. 72, no. 3, id. 034607.

\noindent
\textit{Desilets, D. and Zreda, M.}, Spatial and temporal distribution of secondary cosmic-ray nucleon intensities and applications to in situ cosmogenic dating, Earth Planet. Sci. Lett., 2003, vol. 206, pp. 21-42.

\noindent
\textit{Dunai, T.J.}, Cosmogenic Nuclides. Principles, Concepts and Applications in the Earth Surface Sciences, New York: Cambridge University Press, 2010.

\noindent
\textit{Dunai, T.J., Stuart, F.M., Pik, R., Burnard, P., and Gayer, E.}, Production of $^3$He in crustal rocks by cosmogenic thermal neutrons, Earth Planet. Sci. Lett., 2007, vol. 258, pp. 228-236.

\noindent
\textit{Evenstar, L.A., Hartley, A.J., Stuart, F.M., et al.}, Multiphase development of the Atacama Planation Surface recorded by cosmogenic $^3$He exposure ages: implications for uplift and Cenozoic climate change in western South America, Geology, 2009, vol. 37, no. 1, pp. 27-30.

\noindent
\textit{Farley, K.A., Libarkin, J., Mukhopadhyay, S., and Amidon, W.}, Cosmogenic and nucleogenic $^3$He in apatite, titanite, and zircon, Earth Planet. Sci. Lett., 2006, vol. 248, pp. 451-461.

\noindent
\textit{Feige, Y., Oltman, B.G., and Kastner, J.}, Production rates of neutrons in soils due to natural radioactivity, J. Geophys. Res., 1968, vol. 73, no. 10, pp. 3135-3142.

\noindent
\textit{Folger, G., Ivanchenko, V.N., and Wellisch, J.P.}, The binary cascade, Eur. Phys. J. A, 2004, vol. 21, no. 3, pp. 407-417.

\noindent
\textit{Gayer, E., Pik, R., Lav\'{e}, J., et al.}, Cosmogenic $^3$He in Himalayan garnets indicating an altitude dependence of the $^3$He/$^{10}$Be production ratio, Earth Planet. Sci. Lett., 2004, vol. 229, pp. 91-104.

\noindent
\textit{Goehring, B.M., Kurz, M.D., Balco, G., et al.}, A reevaluation of in situ cosmogenic $^3$He production rates, Quat. Geochronol., 2010, vol. 5, pp. 410-418.

\noindent
\textit{Gordon, M.S., Goldhagen, P., Rodbell, K.P., et al.}, Measurement of the flux and energy spectrum of cosmic-ray induced neutrons on the ground, IEEE Trans. Nucl. Sci.,
2004, vol. 51, no. 6, pp. 3427-3434

\noindent
\textit{Gosse, J.C. and Phillips, F.M.}, Terrestrial in situ cosmogenic nuclides: theory and application, Quat. Sci. Rev., 2001, vol. 20, pp. 1475-1560.

\noindent
\textit{Heikkinen, A., Stepanov, N., and Wellisch, J.P.}, Bertini intra-nuclear cascade implementation in Geant4, ArXiv Nuclear Theory e-prints, 2003, nucl-th/0306008.

\noindent
\textit{Heisinger, B., Lal, D., Jull, A.J.T., et al.}, Production of selected cosmogenic radionuclides by muons 1. Fast muons, Earth Planet. Sci. Lett., 2002a, vol. 200, pp. 345-355.

\noindent
\textit{Heisinger, B., Lal, D., Jull, A.J.T., et al.}, Production of selected cosmogenic radionuclides by muons 2. Capture of negative muons, Earth Planet. Sci. Lett., 2002b, vol. 200, pp. 357-369.

\noindent
\textit{Kober, F., Ivy-Ochs, S., Leya, I., et al.}, In situ cosmogenic $^{10}$Be and $^{21}$Ne in sanidine and in situ cosmogenic $^3$He in Fe-Ti-oxide minerals, Earth Planet. Sci. Lett., 2005, vol. 236, pp. 404-418.

\noindent
\textit{Kurz, M.D.}, Cosmogenic helium in a terrestrial igneous rock, Nature, 1986, vol. 320, pp. 435-439.

\noindent
\textit{Lal, D.}, Cosmic ray labeling of erosion surfaces: in situ nuclide production rates and erosion models, Earth Planet. Sci. Lett., 1991, vol. 104, pp. 424-439.

\noindent
\textit{Lal, D.}, Production of $^3$He in terrestrial rocks, Chem. Geol., 1987, vol. 66, pp. 89-98.

\noindent
\textit{Leya, I., Busemann, H., Baur, H., et al.}, Cross sections for the proton-induced production of He and Ne isotopes from magnesium, aluminum and silicon, Nucl. Instr. Methods Phys. Res. B, 1998, vol. 145, no. 3, pp. 449-458.

\noindent
\textit{Leya, I., Begemann, F., Weber, H.W., et al.}, Simulation of the interaction of galactic cosmic ray protons with meteoroids: on the production of $^3$H and light noble gas isotopes in isotropically irradiated thick gabbro and iron targets, Meteor. Planet. Sci., 2004, vol. 39, no. 3, pp. 367-386.

\noindent
\textit{Licciardi, J.M., Kurz, M.D., Clark, P.U., and Brook, E.J.}, Calibration of cosmogenic $^3$He production rates from Holocene lava flows in Oregon, USA, and effects of the Earth's magnetic field, Earth Planet. Sci. Lett., 1999, vol. 172, pp. 261-271.

\noindent
\textit{Licciardi, J.M., Kurz, M.D., and Curtice, J.M.}, Cosmogenic $^3$He production rates from Holocene lava flows in Iceland, Earth Planet. Sci. Lett., 2006, vol. 246, pp. 251-264.

\noindent
\textit{Lifton, N., Sato, T., and Dunai, T.J.}, Scaling in situ cosmogenic nuclide production rates using analytical approximations to atmospheric cosmic-ray fluxes, Earth Planet. Sci. Lett., 2014, vol. 386, pp. 149-160.

\noindent
\textit{Lippolt, H.J., Wernicke, R.S., and Boschmann, W.}, $^4$He diffusion in specular hematite, Phys. Chem. Miner., 1993, vol. 20, no. 6, pp. 415-418.

\noindent
\textit{Mamyrin, B.A. and Tolstikhin, I.N.}, Izotopy geliya v prirode (Helium Isotopes in Nature), Moscow: Energoizdat, 1981.

\noindent
\textit{Margerison, H.R., Phillips, W.M., Stuart, F.M., and Sugden, D.E.}, Cosmogenic $^3$He concentrations in ancient flood deposits from the Coombs Hills, northern Dry Valleys, East Antarctica: interpreting exposure ages and erosion rates, Earth Planet. Sci. Lett., 2005, vol. 230, pp. 163-175.

\noindent
\textit{Martel, D.J., O'Nions, R.K., Hilton, D.R., and Oxburgh, E.R.}, The role of element distribution in production and release of radiogenic helium: the Carnmenellis Granite, southwest England, Chem. Geol., 1990, vol. 88, pp. 207-221.

\noindent
\textit{Masarik, J. and Reedy, R.C.}, Terrestrial cosmogenic-nuclide production systematics calculated from numerical simulations, Earth Planet. Sci. Lett., 1995, vol. 136, pp. 381-395.

\noindent
\textit{Masarik, J. and Reedy, R.C.}, Monte Carlo simulation of in-situ-produced cosmogenic nuclides, Radiocarbon, 1996, vol. 38, no. 1, pp. 163-164.

\noindent
\textit{Michel, R., Peiffer, F., Theis, S., et al.}, Production of stable and radioactive nuclides in thick stony targets (R = 15 and 25 cm) isotropically irradiated with 600 MeV protons and simulation of cosmogenic nuclides in meteorites, Nucl. Instrum. Methods Phys. Res. B, 1989, vol. 42, no. 1, pp. 76-100.

\noindent
\textit{Michel, R., Gloris, M., Lange, H.-J., et al.}, Nuclide production by proton-induced reactions on elements (6 $\leq$ Z $\leq$ 29) in the energy range from 800 to 2600 MeV, Nucl. Instrum. Methods Phys. Res. B, 1995, vol. 103, pp. 183-222.

\noindent
\textit{Morrison, P. and Pine, J.}, Radiogenic origin of the helium isotopes in rock, Ann. N. Y. Acad. Sci., 1955, vol. 62, no. 3, pp. 71-92.

\noindent
\textit{Nesterenok, A.V.}, Numerical calculations of cosmic ray cascade in the Earth’s atmosphere -- results for nucleon spectra, Nucl. Instrum. Methods Phys. Res. B, 2013, vol. 295, pp. 99-106.

\noindent
\textit{Nesterenok, A.V. and Naidenov, V.O.}, Calculation of the solar activity effect on the production rate of cosmogenic radiocarbon in polar ice, Geomagn. Aeron., 2012a, vol. 52, no. 8, pp. 992-998.

\noindent
\textit{Nesterenok, A.V. and Naidenov, V.O.}, In situ formation of cosmogenic $^{14}$C by cosmic ray nucleons in polar ice, Nucl. Instrum. Methods Phys. Res. B, 2012b, vol. 270, pp. 12-18.

\noindent
\textit{Niedermann, S.}, Cosmic-ray-produced noble gases in terrestrial rocks: dating tools for surface processes, Rev. Mineral. Geochem., 2002, vol. 47, pp. 731-784.

\noindent
\textit{Phillips, F.M., Stone, W.D., and Fabryka-Martin, J.T.}, An improved approach to calculating low-energy cosmic-ray neutron fluxes near the land/atmosphere interface, Chem. Geol., 2001, vol. 175, pp. 689-701.

\noindent
\textit{Porcelli, D., Ballentine, C.J., and Wieler, R.}, An overview of noble gas geochemistry and cosmochemistry, Rev. Mineral. Geochem., 2002, vol. 47, pp. 1-19.

\noindent
\textit{Reedy, R.C.}, Cosmogenic-nuclide production rates: reaction cross section update, Nucl. Instrum. Methods Phys. Res. B, 2013, vol. 294, pp. 470-474.

\noindent
\textit{Rudnick, R.L. and Gao, S.}, Composition of the continental crust, Treat. Geochem., 2003, vol. 3, pp. 1-64.

\noindent
\textit{Sato, T., Yasuda, H., Niita, K., et al.}, Development of PARMA: PHITS-based analytical radiation model in the atmosphere, Radiat. Res., 2008, vol. 170, pp. 244-259.

\noindent
\textit{Shukolyukov, Yu.A.}, Delenie yader urana v prirode (Uranium Fission in Nature), Moscow: Atomizdat, 1970.

\noindent
\textit{Shuster, D.L. and Farley, K.A.}, Diffusion kinetics of proton-induced $^{21}$Ne, $^3$He, and $^4$He in quartz, Geochim. Cosmochim. Acta, 2005, vol. 69, no. 9, pp. 2349-2359.

\noindent
\textit{Starik, I.E.}, Yadernaya geokhronologiya (Nuclear Geochronology), Moscow: AN SSSR, 1961.

\noindent
\textit{Teng, F.-Z., McDonough, W.F., Rudnick, R.L., et al.}, Lithium isotopic composition and concentration of the upper continental crust, Geochim. Cosmochim. Acta, 2004, vol. 68, no. 20, pp. 4167-4178.

\noindent
\textit{Tolstikhin, I.N., Lehmann, B.E., Loosli, H.H., et al.}, Radiogenic helium isotope fractionation: the role of tritium as $^3$He precursor in geochemical applications, Geochim. Cosmochim. Acta, 1999, vol. 63, no. 10, pp. 1605-1611.

\noindent
\textit{Tolstikhin, I.N., Prasolov, E.M., Khabarin, L.V., and Azbel’, I.Ya.}, Geokhimiya radioaktivnykh i radiogennykh izotopov (Geochemistry of Radioactive and Radiogenic Isotopes), Leningrad: Nauka, 1974.

\noindent
\textit{Wieler, R.}, Cosmic-ray-produced noble gases in meteorites, Rev. Mineral. Geochem., 2002, vol. 47, pp. 125-170.

\noindent
\textit{Ziegler, J.F., Ziegler, M.D., and Biersack, J.P.}, SRIM-the stopping and range of ions in matter (2010), Nucl. Instrum. Methods Phys. Res. B, 2010 vol. 268, pp. 1818-1823.

\vspace{1cm}
\noindent
Translated by E. Kurdyukov

\section{Erratum}

There is a mistake in the calculations of nuclide production rates by nucleons and fast muons by Nesterenok A.V., Yakubovich O.V., Petrology, 2016, Vol. 24, pp. 21-34. The calculated production rates of $^3$He by nucleonic component shown in their Table must be divided by rock density 2.65 g/cm$^2$. The same applies to the production rate by fast muons. The production rate by protons was overestimated and constitutes about 10\% after correction. 

The errors mentioned above were fixed in the arxiv version of the manuscript: correct values of the production rate are presented in the Table and figures 4 and 6.

\end{document}